\documentclass[twocolumn,aps,pra,showpacs]{revtex4}

\usepackage{graphicx}

\begin{document}

\title{Bell's theorem with and without inequalities for the
three-qubit Greenberger-Horne-Zeilinger and $W$ states}
\author{Ad\'{a}n Cabello}
\email{adan@us.es}
\affiliation{Departamento de F\'{\i}sica
Aplicada II, Universidad de Sevilla, 41012 Sevilla, Spain}
\date{\today}


\begin{abstract}
A proof of Bell's theorem without inequalities valid for both
inequivalent classes of three-qubit entangled states under local
operations assisted by classical communication, namely
Greenberger-Horne-Zeilinger (GHZ) and
$W$, is described. This proof leads to a Bell inequality that
allows more conclusive tests of Bell's theorem for three-qubit
systems. Another Bell inequality involving both tri- and bipartite
correlations is introduced which illustrates the different
violations of local realism exhibited by the GHZ and $W$ states.
\end{abstract}


\pacs{03.65.Ud,
03.65.Ta}

\maketitle


\section{Introduction}

In recent years, Greenberger-Horne-Zeilinger (GHZ) states of three
or more qubits \cite{GHZ89} have become ubiquitous in quantum
information theory
\cite{CB97,ZZHW98,BVK98,KB98,HBB99,ZLWG00,HLG01}. However, the
interest in GHZ states began in connection with Bell's theorem.
While Bell's proof of the impossibility of Einstein, Podolsky, and
Rosen's (EPR's) ``elements of reality'' \cite{EPR35} was based on
statistical predictions and inequalities \cite{Bell64}, GHZ showed
that a simpler proof can be achieved with perfect correlations and
without inequalities \cite{GHZ89,Mermin90ab,GHSZ90}. Subsequently,
the GHZ proof was translated into experimentally verifiable Bell
inequalities \cite{Mermin90c,RS91} and into real experiments
\cite{BPDWZ99,PBDWZ00}. It has recently been found that not only
the GHZ state but any two-qubit pure entangled state admits a
proof of Bell's theorem without inequalities
\cite{Hardy93,Cabello01b,Cabello01c}.

On the other hand, over the last few years the importance of
quantum entanglement as a resource for unusual kinds of
communication and information processing has stimulated the
mathematical study of the entanglement of multiqubit systems. In
particular, there has been much interest in the classification of
three-qubit pure entangled states in terms of equivalences under
local operations assisted by classical communication (LOCC)
\cite{AACJLT00,DVC00,HS00,CS00,BC01}. D\"{u}r, Vidal, and Cirac
\cite{DVC00} have shown that there are only two classes of
genuinely tripartite entanglement which are inequivalent under
LOCC. One class is represented by the GHZ state,
\begin{equation}
\left| {\rm GHZ} \right\rangle = {1 \over {\sqrt{2}}}
\left( {\left| y+y+y+ \right\rangle + \left| y-y-y-
\right\rangle } \right), \label{GHZstate}
\end{equation}
where $\sigma_y \left| y\pm \right\rangle = \pm \left| y\pm
\right\rangle$, $\sigma_y$ being the corresponding Pauli spin
matrix.
%
%
The other class is represented by the $W$ state \cite{Wolfgang},
\begin{equation}
\left| {W} \right\rangle = {1 \over {\sqrt{3}}}
\left( \left| +-- \right\rangle + \left| -+- \right\rangle +
\left| --+ \right\rangle \right), \label{Wstate}
\end{equation}
where $\sigma_z \left| \pm \right\rangle = \pm \left| \pm
\right\rangle$. The GHZ and $W$ states cannot be converted into
each other by means of LOCC.


At this point, some natural questions arise. The first being in
which applications does the use of the $W$ state mean an
improvement over previous protocols using the GHZ state. This
question is partially addressed in \cite{DVC00}, where it is
pointed out that in a three-qubit system prepared in a $W$ (GHZ)
state, if one of the qubits is traced out then the remaining two
qubits are entangled (completely unentangled). Indeed, $W$ is the
three-qubit state whose entanglement has the highest robustness
against the loss of one qubit \cite{DVC00}. Moreover, from a
single copy of the reduced density matrix for any two qubits
belonging to a three-qubit $W$ state, one can always obtain a
state which is arbitrarily close to a Bell state by means of a
filtering measurement \cite{Gisin96}. This means that, if one of
the parties sharing the system prepared in a $W$ (GHZ) state
decides not to cooperate with the other two, or if for some reason
the information about one of the qubits is lost, then the
remaining two parties still can (cannot) use entanglement
resources to perform communication tasks.

On the other hand, it has been shown that the $W$ state does not
allow a GHZ-type proof of Bell's theorem \cite{Cabello01}.
Therefore, two other natural questions are whether the $W$ state
admits any kind of proof of Bell's theorem without inequalities
and what the differences are between the violation of local
realism exhibited by the GHZ and $W$ states. In this paper, I
will describe four related results which answer these questions.
First, a proof of Bell's theorem without inequalities specific
for the $W$ state. Second, an extension of that proof which is
also valid for the GHZ state. Such a proof leads to a
Bell-type inequality for three qubits which can be experimentally
useful in order to achieve more conclusive tests of Bell's
theorem. Finally, a different set of Bell inequalities is
considered with the purpose of illustrating some differences
between the violations of local realism exhibited by the GHZ and
$W$ states.


\section{\label{sec2}Bell's theorem without inequalities for the $W$ state}

First, I will show that the $W$ state allows three local
observers to define elements of reality which are incompatible
with some predictions of quantum mechanics \cite{Pitowsky91}. I
will use the following notation: $z_q$ and $x_q$ will be the
results ($-1$ or $1$) of measuring $\sigma_z$ and $\sigma_x$ on
qubit $q$ ($q=1,2,3$). The first step of the proof consists of
showing that, in the $W$ state, all the $z_q$ and $x_q$ satisfy
EPR's criterion of elements of reality. EPR's condition reads:
{\em ``If, without in any way disturbing a system, we can predict
with certainty (i.e., with probability equal to unity) the value
of a physical quantity, then there exists an element of physical
reality corresponding to this physical quantity''} \cite{EPR35}.
From the expression of the $W$ state given in Eq.~(\ref{Wstate}), it
can be immediately seen that $z_1$, $z_2$, and $z_3$ are elements
of reality, since the result $z_i$ can be predicted with certainty
from the results of spacelike separated measurements of $z_j$ and
$z_k$ ($i \neq j$, $j \neq k$, and $k \neq i$). In addition, by
rewriting the $W$ state in the suitable basis, it can easily be
seen that, if $z_i=-1$ then, with certainty, $x_j=x_k$. Therefore,
if $z_i=-1$, then by measuring $x_j$ ($x_k$) one can predict $x_k$
($x_j$) with certainty. Therefore, if $z_i=-1$, then $x_j$ and
$x_k$ are elements of reality. If $z_i=+1$ then, using the
expression~(\ref{Wstate}), it can immediately be seen that
$z_j=-1$. Therefore, following the previous reasoning, $x_i$ and
$x_k$ are elements of reality (although $x_i$ could have ceased to
be an element of reality after measuring $\sigma_z$ on particle
$i$). In conclusion, $z_q$ and $x_q$ are EPR elements of reality
and therefore, according to EPR, they should have predefined
values $-1$ or $1$ before any measurement.

However, according to quantum mechanics, such an assignment of
values is impossible. The proof can be presented in a very similar
way to Hardy's proof of Bell's theorem for nonmaximally entangled
states of two qubits \cite{Hardy93} by using four properties of
the quantum state and a logical argument based on them. For the
refutation of EPR's elements of reality, the relevant properties
of the $W$ state~(\ref{Wstate}), which can easily be checked, are
\begin{eqnarray}
P_W \left( {z_i=-1,z_j=-1} \right)=1,
\label{one} \\
P_W \left( {x_j=x_k} | {z_i=-1} \right)=1,
\label{two} \\
P_W \left( {x_i=x_k} | {z_j=-1} \right)=1,
\label{three} \\
P_W \left( {x_i=x_j=x_k} \right)={3 \over 4},
\label{four}
\end{eqnarray}
where $P_W \left( {z_i=-1,z_j=-1} \right)$ means the probability
of two qubits (although we cannot tell which one) giving the result
$-1$ when measuring $\sigma_{z}$ on all three qubits, and $P_W
\left( {x_j=x_k} | {z_i=-1} \right)$ is the conditional
probability of $\sigma_{xj}$ and $\sigma_{xk}$ having the same
result given that the result of $\sigma_{zi}$ is $-1$.
Property~(\ref{one}) tells us that, when measuring
$\sigma_z$ on all three
qubits, the result $-1$ {\em always} occurs in {\em two} of them.
Let us call these qubits $i$ and $j$. Then let us suppose that we
had measured $\sigma_x$ on qubits $j$ and $k$, instead of
$\sigma_z$. Then, according to property~(\ref{two}), the results
would have been the same. Therefore, following EPR, one reaches
the conclusion that the predefined values corresponding to the
elements of reality $x_j$ and $x_k$ are equal. Now let us suppose
that we had measured $\sigma_x$ on qubits $i$ and $k$, instead of
$\sigma_z$. Then, according to property~(\ref{three}), the results
would have been the same. Therefore, the predefined values
corresponding to the elements of reality $x_i$ and $x_k$ are
equal. Taking these two conclusions together, one must deduce
that, in the $W$ state, the predefined values corresponding to the
elements of reality $x_1$, $x_2$, and $x_3$ {\em always} satisfy
$x_1=x_2=x_3$. However, this is in contradiction with
property~(\ref{four}) which states that,
when measuring $\sigma_x$ on all
three qubits, one finds results that {\em cannot} be explained
with elements of reality in ${1 \over 4}$ of the cases. Therefore,
the conclusion is that quantum predictions for the $W$ state
cannot be ``completed'' with EPR's elements of reality.


\begin{figure}
\includegraphics[width=5.5cm]{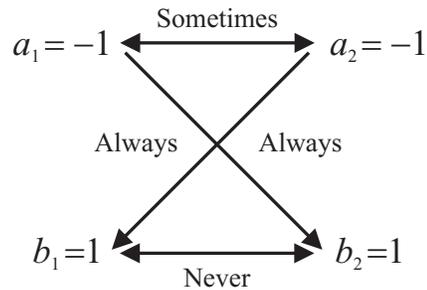}
\caption{\label{fig2}
Diagram for Hardy's proof of Bell's
theorem without inequalities for two qubits in a nonmaximally
entangled state \cite{Hardy93,BBMH97}. $a_i$ and $b_i$ are
alternative spin observables of qubit $i$.}
\end{figure}

\begin{figure}
\includegraphics[width=5.5cm]{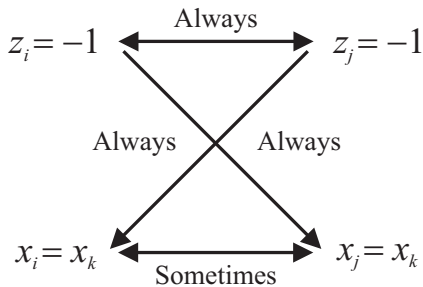}
\caption{\label{fig3}
Diagram for the proof of Bell's theorem
without inequalities for the $W$ state.}
\end{figure}

While the structure of this proof is similar to Hardy's
\cite{Hardy93,BBMH97}, the logical argument is different: in
Hardy's, from a result that {\em sometimes} occurs, two local
observers infer a result that {\em never} occurs (see Fig.~\ref{fig2});
here, from a result that {\em always} occurs, two observers infer
a result that {\em only sometimes} occurs (see Fig.~\ref{fig3}). In
addition, while in Hardy's proof only 9\% of the runs of a certain
experiment contradict local realism, here 25\% of the runs of the
last experiment cannot be explained by local realism. On the other
hand, while in Hardy's proof we need both qubits to start the
argument and in GHZ's proof \cite{Mermin90ab,GHSZ90} all three
qubits are required, in the proof for the $W$ state the
contradiction results from inferences from only {\em two} of all
three qubits, but we cannot tell which one.


\section{\label{sec3}Bell's theorem without inequalities
for the GHZ and $W$ states}

\begin{figure}
\includegraphics[width=5.5cm]{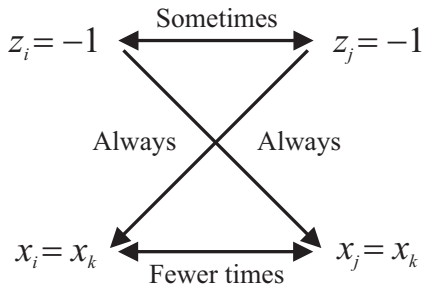}
\caption{\label{fig4}
Diagram for the extended proof of Bell's
theorem without inequalities valid for the GHZ and $W$ states. The
diagram in Fig.~\ref{fig3} is a particular case of this one.}
\end{figure}

For two-qubit pure states, the logical structure that can be
obtained from Fig.~\ref{fig2} by changing
the ``never'' to ``fewer times''
is not particularly useful, since the states which satisfy a
``sometimes-always-never'' structure
like that in Fig.~\ref{fig2}, namely,
nonmaximally entangled states \cite{Hardy93}, are the same which
satisfy the extended structure \cite{Lucien}. However, for
three-qubit pure states, a similar extension of Fig.~\ref{fig3}'s
``always-always-sometimes'' structure (changing the first
``always'' to ``sometimes'' and the last ``sometimes'' to ``fewer
times'') allows us to extend the proof for the $W$ state to the
GHZ state and, therefore, to obtain a common proof of Bell's
theorem without inequalities for both classes of genuinely
tripartite entangled states. Such a proof
is illustrated in Fig.~\ref{fig4}
and, for the GHZ state~(\ref{GHZstate}), it is based on its
following four properties which can easily be checked:
\begin{eqnarray}
P_{\rm GHZ} \left( {z_i=-1,z_j=-1} \right)={3 \over 4},
\label{GHZone} \\
P_{\rm GHZ} \left( {x_j=x_k} | {z_i=-1} \right)=1,
\label{GHZtwo} \\
P_{\rm GHZ} \left( {x_i=x_k} | {z_j=-1} \right)=1,
\label{GHZthree} \\
P_{\rm GHZ} \left( {x_i=x_j=x_k} \right)= {1 \over 4}.
\label{GHZfour}
\end{eqnarray}
The proof for the GHZ state is parallel to the one for the $W$
state, changing only the ``always'' to ``in 75\% of the cases''
and changing the percentage of events of the fourth experiment
that cannot be explained with elements of reality, which now is
50\%.


The previous demonstrations complete the family of simple proofs
of Bell's theorem without inequalities for the main classes of
two- \cite{Hardy93,Cabello01b,Cabello01c} and three-qubit
\cite{GHZ89,Mermin90ab,GHSZ90} pure entangled states.


\section{Bell-CH inequalities for three qubits}

The proofs without inequalities described
in Secs.~\ref{sec2} and~\ref{sec3} can easily
be converted into experimentally testable Bell inequalities. As
noted in \cite{Hardy94,Mermin94,Mermin95}, the two-qubit Bell
inequalities proposed by Clauser and Horne (CH) \cite{CH74} can be
put into a form in which only the four probabilities of a
Hardy-type argument are included. Similarly, it can easily be seen
that four probabilities related to those
in Eqs.~(\ref{one})--(\ref{four}) [or in Eqs.~(\ref{GHZone})--(\ref{GHZfour})]
must satisfy the following Bell inequality:
\begin{eqnarray}
\! \! \! \!
\! \! \! \! \! \! -1 \! \! & \le & \! \! P \left( {z_i=-1,z_j=-1} \right) -
P \left( {z_i=-1,x_j \neq x_k} \right)
\nonumber \\
\! \! \! \! \! \! \! \! & & \! \!
\! \!
-P \left( {x_i \neq x_k, z_j=-1} \right)-
 P \left( {x_i=x_j=x_k} \right) \le 0.
\label{inequalities}
\end{eqnarray}
For the $W$ state~(\ref{Wstate}), the value of the first
probability in Eq.~(\ref{inequalities}) is 1, the values of the second
and third probabilities are 0, and the value of the fourth is 3/4.
Therefore, the middle term in Eq.~(\ref{inequalities}) is 0.25. This
means that the violation of the inequality~(\ref{inequalities}) is
higher than the maximum violation obtained from Hardy's proof,
where the middle term is 0.090 \cite{Mermin94,Mermin95}, and even
higher than the violation for a singlet state, where the middle
term is 0.207 \cite{Mermin94,Mermin95}. The violation exhibited by
the GHZ state is even higher.
By using properties~(\ref{GHZone})--(\ref{GHZfour}),
it can easily be seen that the
middle term in Eq.~(\ref{inequalities}) is 0.5, which is the maximum
allowed violation of a CH-type inequality and corresponds to a
value four in the corresponding Clauser-Horne-Shimony-Holt (CHSH)
inequality \cite{CHSH69}.


A similar situation occurs in Mermin's inequality \cite{Mermin90c}
\begin{eqnarray}
-2 \! \! & \le & \! \!
\langle A_1 A_2 A_3 \rangle -
\langle A_1 B_2 B_3 \rangle -
\langle B_1 A_2 B_3 \rangle \nonumber \\
\! \! & & \! \!
-\langle B_1 B_2 A_3 \rangle \le 2,
\label{Mermininequality}
\end{eqnarray}
where $A_i$ and $B_i$ are observables of qubit $i$. By choosing
$A_i=\sigma_{zi}$ and $B_i=\sigma_{xi}$,
for the GHZ state~(\ref{GHZstate}) we obtain four for the
middle term in Eq.~(\ref{Mermininequality}),
four being the maximum allowed
violation of inequality~(\ref{Mermininequality}).
For the $W$ state~(\ref{Wstate}), considering
only local spin observables on plane $x$-$z$, and that
$A_1=A_2=A_3$ and $B_1=B_2=B_3$, the maximum violation is 3.046
[for instance, by choosing $A_i=\cos(0.628) \sigma_{x} -
\sin(0.628) \sigma_{z}$ and $B_i=\cos(1.154) \sigma_{x} +
\sin(1.154) \sigma_{z}$]. Alternatively, by choosing
$A_i=\sigma_{zi}$ and $B_i=\sigma_{xi}$ (which satisfy EPR's
criterion of elements of reality), the $W$ state~(\ref{Wstate})
gives the value three for the middle term in Eq.~(\ref{Mermininequality}).

Two reasons suggest that, for the three-qubit GHZ state,
inequality~(\ref{inequalities}) could lead to a more conclusive
clear-cut experimental test of Bell's theorem
than inequality~(\ref{Mermininequality}).
As in CH's, inequality~(\ref{inequalities})
can be put into a form which does not involve
the number of undetected particles, thereby rendering unnecessary
the assumption of fair sampling \cite{CH74}. On the other hand,
since CH and CHSH inequalities are equivalent \cite{Mermin95},
and Eq.~(\ref{inequalities}) and CH [Eq.~(\ref{Mermininequality}) and CHSH]
inequalities have the same bounds, the ratio between the maximum
violations shown by the GHZ and singlet states
for inequality~(\ref{inequalities}) over CH's
[Eq.~(\ref{Mermininequality}) over
CHSH's], $1+\sqrt{2}$ [$\sqrt{2}$], suggests that
Eq.~(\ref{inequalities}) reveals a higher violation of local realism
than Eq.~(\ref{Mermininequality}).


\section{Bell inequalities involving tri- and bipartite correlations}

As a final remark, the results in \cite{DVC00} point out that
bipartite correlations are relevant to the $W$ state but not to
the GHZ state. Therefore, it would be interesting to consider Bell
inequalities involving both tripartite and bipartite correlations.
The simplest way of obtaining such an inequality would be by
adding genuinely bipartite correlations to the tripartite
correlations considered in Mermin's inequality.
For instance, a straightforward calculation would allow us to
prove that any local realistic theory must satisfy the following
inequality:
\begin{eqnarray}
-5 \! \! & \le & \! \!
\langle A_1 A_2 A_3 \rangle \! - \!
\langle A_1 B_2 B_3 \rangle \! - \!
\langle B_1 A_2 B_3 \rangle \! - \!
\langle B_1 B_2 A_3 \rangle \!
\nonumber \\
\! \! & & \! \!
-\langle A_1 A_2 \rangle \! - \!
\langle A_1 A_3 \rangle \! - \!
\langle A_2 A_3 \rangle \le 3.
\label{Bell3plus2}
\end{eqnarray}
Assuming that $A_i$ and $B_i$ are local observables on plane
$x$-$z$, and that $A_1=A_2=A_3$ and $B_1=B_2=B_3$, a numerical
calculation shows that both the GHZ and $W$ states give a maximum
value four for the middle term in Eq.~(\ref{Bell3plus2}) (for instance,
by choosing $A_i=\sigma_{zi}$ and $B_i=\sigma_{xi}$ in both
cases). Therefore, both states lead to the {\em same} maximal
violation of the inequality~(\ref{Bell3plus2}). However, if we
assign a higher weight to the bipartite correlations appearing in
the inequality, then we can reach a Bell inequality such as
\begin{eqnarray}
-8 \! \! & \le & \! \!
\langle A_1 A_2 A_3 \rangle \! - \!
\langle A_1 B_2 B_3 \rangle \! - \!
\langle B_1 A_2 B_3 \rangle \! - \!
\langle B_1 B_2 A_3 \rangle \! \nonumber \\
\! \! & & \! \!
-2\, \langle A_1 A_2 \rangle \! - \!
2\, \langle A_1 A_3 \rangle \! - \!
2\, \langle A_2 A_3 \rangle \le 4,
\label{Bell2plus3}
\end{eqnarray}
which is violated by the $W$ state [for instance, by choosing
$A_i=\sigma_{zi}$ and $B_i=\sigma_{zi}$, state~(\ref{Wstate})
gives the value five for the middle term in Eq.~(\ref{Bell2plus3})] but
{\em not} by the GHZ state. Therefore, there are scenarios
involving both tripartite and bipartite correlations in which the
quantum predictions for the GHZ state can be reproduced with a
local model while those for the $W$ state cannot.
A more general study of multipartite Bell inequalities is
presented in \cite{PS01,WW01}.


\section{Summary}

In brief, we have completed the family of proofs without
inequalities for two- and three-qubit pure entangled states with a
proof for the $W$ state that can also be extended to the GHZ state
and we have then obtained two Bell inequalities for three
qubits. The first could lead to more conclusive tests of Bell's
theorem. The second, involving both tri- and bipartite
correlations, illustrates some differences between the violations
of local realism exhibited by the GHZ and $W$ states.


\begin{acknowledgments}
I thank J. L. Cereceda and C. Serra
for their comments, and the organizers of the
Sixth Benasque Center for Science, where this work was begun, the
Junta de Andaluc\'{\i}a Grant No.~FQM-239, and the Spanish Ministerio
de Ciencia y Tecnolog\'{\i}a Grant No.~BFM2000-0529 for their support.
\end{acknowledgments}



\begin{thebibliography}{99}
\bibitem{Wolfgang} The name ``$W$'' is for Wolfgang
(D\"{u}r), one of the authors of Ref. \cite{DVC00}.
\end{thebibliography}
\end{document}